\documentclass[a4paper]{jpconf}

\usepackage{graphicx}
\usepackage{amsmath,amssymb}
\usepackage{subfigure}
\usepackage{color}
\usepackage{multirow}
\usepackage{cite}

\begin{document}

\title{A frustrated spin-1 $J_{1}$--$J_{2}$ Heisenberg antiferromagnet: An anisotropic planar pyrochlore model}

\author{P H Y Li$^1$, R F Bishop$^1$, 
and C E Campbell$^2$}

\address{$^1$ School of Physics and Astronomy, Schuster Building, The University of Manchester, Manchester, M13 9PL, UK}
\address{$^2$ School of Physics and Astronomy, University of Minnesota, 116 Church Street SE, Minneapolis, Minnesota 55455, USA}

\ead{raymond.bishop@manchester.ac.uk; peggyhyli@gmail.com}

\begin{abstract}
  The zero-temperature ground-state (GS) properties and phase diagram
  of a frustrated spin-1 $J_{1}$--$J_{2}$ Heisenberg model on the
  checkerboard square lattice are studied, using the coupled cluster
  method.  We consider the case where the nearest-neighbour exchange
  bonds have strength $J_{1}>0$ and the next-nearest-neighbour
  exchange bonds present (viz., in the checkerboard pattern of the
  planar pyrochlore) have strength $J_{2} \equiv \kappa J_{1}>0$.  We
  find significant differences from both the spin-1/2 and classical
  versions of the model.  We find that the spin-1 model has a first
  phase transition at $\kappa_{c_{1}} \approx 1.00 \pm 0.01$ (as does
  the classical model at $\kappa_{{\rm cl}}=1$) between two
  antiferromagnetic phases, viz., a quasiclassical N\'{e}el phase (for
  $\kappa < \kappa_{c_{1}}$) and one of the infinitely degenerate
  family of quasiclassical phases (for $\kappa > \kappa_{c_{1}}$) that
  exists in the classical model for $\kappa > \kappa_{{\rm cl}}$,
  which is now chosen by the {\it order by disorder} mechanism as
  (probably) the ``doubled N\'{e}el'' (or N\'{e}el$^{\ast}$) state.
  By contrast, none of this family survives quantum fluctuations to
  form a stable GS phase in the spin-1/2 case.  We also find evidence
  for a second quantum critical point at $\kappa_{c_{2}} \approx 2.0
  \pm 0.5$ in the spin-1 model, such that for $\kappa >
  \kappa_{c_{2}}$ the quasiclassical (N\'{e}el$^{\ast}$) ordering
  melts and a nonclassical phase appears, which, on the basis of
  preliminary evidence, appears unlikely to have crossed-dimer
  valence-bond crystalline (CDVBC) ordering, as in the spin-1/2 case.
  Unlike in the spin-1/2 case, where the N\'{e}el and CDVBC phases are
  separated by a phase with plaquette valence-bond crystalline (PVBC)
  ordering, we find very preliminary evidence for such a PVBC state in
  the spin-1 model for all $\kappa > \kappa_{c_{2}}$.
\end{abstract}

\section{Introduction}
\label{intro}
Low-dimensional spin-lattice models of magnetic systems, particularly
those pertaining to frustrated Heisenberg antiferromagnets (HAFMs) with
competing interactions, have been extensively studied from both the
theoretical and experimental viewpoints in recent years.  Although
such spin-lattice models are themselves conceptually simple and easy
to write down, these strongly correlated systems often exhibit rich
and interesting zero-temperature ($T=0$) ground-state (GS) phase
diagrams as the interaction coupling strengths are varied, due to the
strong interplay between quantum fluctuations and frustration.  The
strength of the quantum fluctuations can itself also be tuned by a
variety of methods.  These include changing the spin quantum number
$s$ of the particles residing on the given lattice sites, while keeping the
interaction Hamiltonian unchanged.  One expects that, in general,
quantum fluctuations will be greatest for the case $s=\frac{1}{2}$,
and that they will reduce to zero as the classical limit ($s
\rightarrow \infty$) is approached.

For this reason the greatest attention has been paid to spin-1/2
magnets.  Nevertheless, there has also been an upsurge of interest in
spin-1 magnets in recent years.  On the theoretical side, although
quantum fluctuations will generally be reduced for a given model for
the case $s=1$ compared with its counterpart for the case
$s=\frac{1}{2}$, totally new physical effects can also sometimes
enter.  In one-dimensional (1D) systems these include the by now
well-known existence of the gapped Haldane phase~\cite{Ha:1983} with an
exponential decay with separation of spin-spin correlations for $s=1$
(or, more generally, for integral values of $s$), compared to the
gapless phase with a corresponding power-law decay of spin-spin
correlations for $s=\frac{1}{2}$ (or, more generally, for
half-odd-integral values of $s$).  Also, in any number of dimensions,
the possible inclusion in the $s=1$ case of such additional terms in
the Hamiltonian as biquadratic exchange and single-site anisotropy,
which are absent for $s=\frac{1}{2}$ systems, can lead both to novel
quantum phase transitions and to novel phases with, for example,
quadrupolar nematic long-range order (LRO) but with zero magnetic
order parameter (taken as the average local on-site magnetization).

On the experimental side many magnetic compounds
containing spin-1 ions are now established as being well described by
various $s=1$ spin-lattice models.  For the 1D case there are many
good experimental realizations of (quasi-)linear chain systems with
$s=1$.  These include CsNiCl$_{3}$~\cite{St:1987_s1} and
Y$_{2}$BaNiO$_{5}$~\cite{Daniel:1993_s1}, both with a weak easy-axis
single-ion anisotropy, and CsFeBr$_{3}$~\cite{Dorner:1988_s1} with a
strong easy-plane single-ion anisotropy, as well as such complex
organo-metallic compounds as NENP
(Ni(C$_{2}$H$_{8}$N$_{2}$)$_{2}$(NO$_{2}$)(ClO$_{4}$))~\cite{Renarnd:1988_s1}
with a weak planar anisotropy, and both NENC
(Ni(C$_{2}$H$_{8}$N$_{2}$)$_{2}$Ni(CN$_{4}$))~\cite{Orendac:1995_s1}
and DTN (NiCl$_{2}$-4SC(NH$_{2}$)$_{2}$)~\cite{Zapf:2006_s1} with a
strong planar anisotropy.  The spin gaps seen in CsNiCl$_{3}$,
Y$_{2}$BaNiO$_{5}$ and NENP are now believed to be experimental
realizations of the integer-spin gap behaviour predicted by
Haldane~\cite{Ha:1983}.  For the two-dimensional (2D) case several experimental
realizations of spin-1 antiferromagnets exist.  For example,
K$_{2}$NiF$_{4}$~\cite{Birgeneau:1970_s1} provides a realization of an
$s=1$ HAFM on a square lattice.  Similarly,
NiGa$_{2}$S$_{4}$~\cite{Nakatsuji:2005_s1} is well described as a 2D
antiferromagnet on a triangular lattice, for which the GS phase has
been argued to have ferro-spin nematic
order~\cite{Lauchli:2006_s1,Tsunetsugu:2006_s1,Bhattacharjee:2006_s1}.

Of particular relevance to the present study has been the large
additional impetus to the study of 2D spin-1 antiferromagnets that was
provided by the discovery of superconductivity with a transition
temperature $T_c \approx 26\,$K in the layered iron-based compound
LaOFeAs, when doped by partial substitution of the oxygen atoms by
fluorine atoms~\cite{KWHH:2008_s1}, La[O$_{1-x}$F$_x$]FeAs, with $x
\approx$ 5--11\%.  That discovery was rapidly followed by finding
superconductivity at even higher transition temperatures ($T_c \gtrsim
50\,$K) in a wide class of similarly doped quaternary oxypnictide
materials.  First-principles calculations~\cite{MLX:2008_s1} ensued,
which showed that the original undoped parent precursor material
LaOFeAs is well described by the spin-1 $J_{1}$--$J_{2}$ HAFM on a
square lattice with nearest-neighbour (NN) and next-nearest-neighbour
(NNN) Heisenberg exchange couplings, $J_{1}$ and $J_{2}$ respectively,
with $J_1 > 0$, $J_2 > 0$, and with $J_{2}/J_{1} \approx 2$.  Other authors
have also reached similar conclusions (see, e.g.,
Ref.~\cite{SA:2008_s1}).

The $J_{1}$--$J_{2}$ model on a square lattice has itself received
huge theoretical attention over the last 25 or so years, since it
provides an archetypal model of a strongly correlated and highly
frustrated spin-lattice system.  Most attention has naturally been
devoted to the spin-1/2 case (see, e.g.,
Refs.~\cite{Chandra:1988,Gelfand:1988,Dagotto:1989,Sachdev:1990,Chubukov:1991,Read:1991,Schulz:1992,Richter:1993,Richter:1994,Oitmaa:1996,Schulz:1996,Zhitomirsky:1996,Bishop:1998_J1J2mod,Singh:1999,Kotov:1999,Capriotti:2000,Capriotti:2001,Takano:2003,Roscilde:2004,Sirker:2006,Mambrini:2006,Schm:2006,Darradi:2008,Ralko:2009,Richter:2010_ED40,Reuther:2011_J1J2J3mod,Yu:2012,Gotze:2012,Jiang:2012,Mezzacapo:2012,LiT:2012,Wang:2013,Hu:2013}
and references therein).  The consensual view for this model now is
that its ($T=0$) GS phase diagram exhibits two phases with
quasiclassical LRO, both with antiferromagnetic (AFM) order, namely, a
N\'{e}el-ordered phase (with a wavevector ${\mathbf Q} = (\pi,\pi)$) at small
values of the frustration parameter ($J_{2}/J_{1} \lesssim 0.4$) and a
collinear stripe-ordered phase (with a wavevector ${\mathbf Q} = (\pi,0)$ or ${\mathbf Q}
= (0,\pi)$) at large values ($J_{2}/J_{1} \gtrsim 0.6)$.  These two
magnetically ordered phases are separated by an intermediate quantum
paramagnetic (QP) phase without magnetic LRO for $0.4 \lesssim
J_{2}/J_{1} \lesssim 0.6$.  What makes the system of continuing
interest is that the nature of the intermediate QP phase and the order
and nature of the two phase transitions bounding it are still not
fully resolved and understood.

The classical ($s \rightarrow \infty$) version of the $J_{1}$--$J_{2}$
model on the square lattice (with a number $N \rightarrow \infty$ of
spins) exhibits a unique GS N\'{e}el-ordered AFM phase for
$J_{2}/J_{1} < \frac{1}{2}$, with an energy per spin $E^{\rm cl}/N =
-2s^{2}(J_{1}-J_{2})$, but has an infinitely degenerate set of GS
phases for $J_{2}/J_{1} > \frac{1}{2}$, all with $E^{\rm cl}/N =
-2s^{2}J_{2}$.  The latter set comprises two interpenetrating
N\'{e}el-ordered $\sqrt{2}\times\sqrt{2}$ square lattices, with the
relative ordering angle between them completely arbitrary.  Quantum
fluctuations then act, via the well known {\it order by disorder}
mechanism \cite{Villain:1977,Shender:1982}, to lift this (accidental)
degeneracy in the quasiclassical ($s \gg 1$) limit where one works to
leading order in $1/s$, in favour of collinear ordering, which leads
to the two (row or column) stripe-ordered states, with wavevectors
${\mathbf Q} = (0,\pi)$ and ${\mathbf Q} = (\pi,0)$, discussed above.

This feature of macroscopic classical GS degeneracy in {\it any}
spin-lattice model always makes such models of particular theoretical
interest since they are, {\it a priori}, prime candidates for
exhibiting novel quantum GS phases.  It also makes them particularly
susceptible to small perturbations in the form, for example, of
additional spin-orbit interactions, spin-lattice couplings, neglected
exchange terms, and anisotropies in the exchange interactions.  Hence,
considerable attention has also been placed on various such mechanisms,
or extra parameters that can be included, to extend the
$J_{1}$--$J_{2}$ model on the square lattice, apart from changing the
spin quantum number, both to learn more about the model itself and to enquire
how robust are its various properties against any such perturbations.

Naturally, for the purpose of making such detailed comparisons, it is
important to use an accurate theoretical technique with controlled
approximation hierarchies.  One such is the coupled cluster method
(CCM) \cite{Bi:1991_TheorChimActa,Bishop:1998,Ze:1998,Fa:2004} that we shall employ
here, and which we discuss in more detail in Sec.~\ref{ccm_section}.
The CCM has been very successfully applied over the last 20 or more
years to a wide variety of quantum spin-lattice systems (see, e.g.,
Refs.~\cite{Bishop:1998_J1J2mod,Schm:2006,Darradi:2008,Reuther:2011_J1J2J3mod,Gotze:2012,Ze:1998,Fa:2004,RoHe:1990,Bishop:1991_PRB43,Bishop:1992_PRB46,Bishop:1992_JPCM4,Bishop:1993_JPCM5,Zeng:1996,Bishop:2000,Kr:2000,Fa:2001_PRB64,Farnell:2002,Darradi:2005_PRB72,Farnell:2008,Bi:2008_PRB_J1xxzJ2xxz,Bi:2008_J1J1primeJ2,Bi:2008_EPL,Bi:2008_JPCM_J1xxzJ2xxz_s1,Farnell:2009,Bi:2009_SqTriangle,Richter:2010:J1J2mod_FM,Bishop:2010_UJack,Bishop:2010_KagomeSq,Bishop:2010_UJack_GrtSpins,DJJF:2011_honeycomb,Gotze:2011_kagome,PHYLi:2011_SqTriangle_grtSpins,PHYLi:2012_honeycomb_J1neg,Bishop:2012_honeyJ1-J2,Bishop:2012_Honeycomb_J2neg,Bishop:2012_checkerboard,Li:2012_honey_full,Li:2012_anisotropic_kagomeSq,RFB:2013_hcomb_SDVBC,Li:2013_chevron,Bishop:2013_cross-striped}
and references cited therein).  These include applications both to the
square-lattice $J_{1}$--$J_{2}$ model itself
\cite{Bishop:1998_J1J2mod,Darradi:2008,Gotze:2012,Richter:2010:J1J2mod_FM} as well as to
various extensions and refinements of it along the lines discussed
above.  

Such extensions include, {\it inter alia}: (a) the
$J_{1}$--$J_{2}$--$J_{\perp}$ model of a stacked square lattice
\cite{Schm:2006}, in which a number of 2D $J_{1}$--$J_{2}$
square-lattice layers are coupled via a NN inter-layer exchange
interaction of strength $J_{\perp}$; (b) putting a square-plaquette
structure on the model \cite{Gotze:2012} by having differing inter-
and intra-plaquette NN couplings; (c) the corresponding
$J_{1}$--$J_{2}$--$J_{3}$ model \cite{Reuther:2011_J1J2J3mod}, which
includes next-next-nearest-neighbour Heisenberg couplings of strength
$J_{3}$; (d) the $J_{1}$--$J_{1}'$--$J_{2}$ model
\cite{Bi:2008_J1J1primeJ2}, in which a spatial anisotropy between NN
bonds along the two perpendicular square-lattice directions is
introduced; and (e) the $J_{1}^{XXZ}$--$J_{2}^{XXZ}$ model
\cite{Bi:2008_PRB_J1xxzJ2xxz}, in which an $XXZ$-type anisotropy is
introduced on both the NN and NNN Heisenberg exchange bonds.  The
corresponding $s=1$ cases have also been studied within the CCM
framework for the latter two cases of the $J_{1}$--$J_{1}'$--$J_{2}$
model \cite{Bi:2008_EPL} and the $J_{1}^{XXZ}$--$J_{2}^{XXZ}$ model
\cite{Bi:2008_JPCM_J1xxzJ2xxz_s1} on the square lattice.

Of particular importance for, and relevance to, the present paper, we
note that the CCM has also been applied to study the ($T=0$) GS phase
diagrams of several members of the so-called half-depleted spin-1/2 $J_{1}$--$J_{2}$
models on the square lattice, all of which share the feature that half
of the $J_{2}$ bonds of the original model are removed.  They differ
only in the arrangements of the remaining $J_{2}$ bonds.  When each
basic square plaquette (formed from 4 NN $J_{1}$ bonds) has a single
$J_{2}$ bond, they include the three cases of: (a) the interpolating
square-triangle model \cite{Bi:2009_SqTriangle}, in which the $J_{2}$
bonds have the same orientation in each square plaquette; (b) the
Union Jack model \cite{Bishop:2010_UJack}, in which the $J_{2}$ bonds
have alternating orientations on neighbouring square plaquettes; and
(c) the chevron-decorated square-lattice model \cite{Li:2013_chevron},
in which the $J_{2}$ bonds alternate in orientation in one direction
(say, along rows), but are parallel to each other in the perpendicular
direction (say, along columns).

The corresponding $s=\frac{1}{2}$ model on the checkerboard lattice,
in which alternating basic square plaquettes have either both or zero
$J_{2}$ bonds present, has also been studied within the CCM
\cite{Bishop:2012_checkerboard}.  Furthermore, CCM studies have also
been carried out for $s \geq 1$ cases of both the interpolating
square-triangle lattice model \cite{PHYLi:2011_SqTriangle_grtSpins}
and the Union Jack model \cite{Bishop:2010_UJack_GrtSpins}, which both show
interesting differences to their $s=\frac{1}{2}$ counterparts.  The
aim of the present work is to perform a similar CCM analysis of the
$s=1$ version of the $J_{1}$--$J_{2}$ model on the checkerboard
lattice, and to compare it with its $s=\frac{1}{2}$ counterpart
studied previously by us \cite{Bishop:2012_checkerboard}.

The $J_{1}$--$J_{2}$ model on the checkerboard lattice, shown
schematically in Fig.~\ref{model_bonds}, is also known as the
anisotropic planar pyrochlore (APP) model (or, sometimes, the crossed
chain model).  It may be regarded as a 2D analogue of a
three-dimensional (3D) anisotropic pyrochlore model of corner-sharing
tetrahedra.  The model itself, as we elaborate in
Sec.~\ref{model_sec}, falls into the same interesting class as the
full $J_{1}$--$J_{2}$ model on the square lattice, demonstrating
macroscopic classical GS degeneracy above a certain critical value of
the anisotropy parameter, $\kappa \equiv J_{2}/J_{1}$.  For that
reason the ($T=0$) GS phase diagram of the $s=\frac{1}{2}$ version of
the model has been much studied earlier by many authors
\cite{Bishop:2012_checkerboard,Singh:1998,Palmer:2001,Brenig:2002,Canals:2002,Starykh:2002,Sindzingre:2002,Fouet:2003,Berg:2003,Tchernyshyov:2003,Moessner:2004,Hermele:2004,Brenig:2004,Bernier:2004,Starykh:2005,Schmidt:2006,Arlego:2007,Moukouri:2008,Chan:2011}.
More recently, on the basis of our CCM results for this model
\cite{Bishop:2012_checkerboard}, a more accurate and, hopefully, more
consensual description of its GS phase structure is emerging.  The
time now seems ripe, therefore, to compare and contrast the model with
its $s=1$ counterpart, which we study here.

In Sec.~\ref{model_sec} the model itself is discussed, before we give
a brief description of the CCM formalism in Sec.~\ref{ccm_section}.
Our main results are then presented in Sec.~\ref{results}, and we end
with a discussion and summary in Sec.~\ref{summary}.  

\section{The model}
\label{model_sec}
\begin{figure}[!tb]
\begin{center}
\mbox{
\subfigure[N\'{e}el]{\scalebox{0.4}{\includegraphics{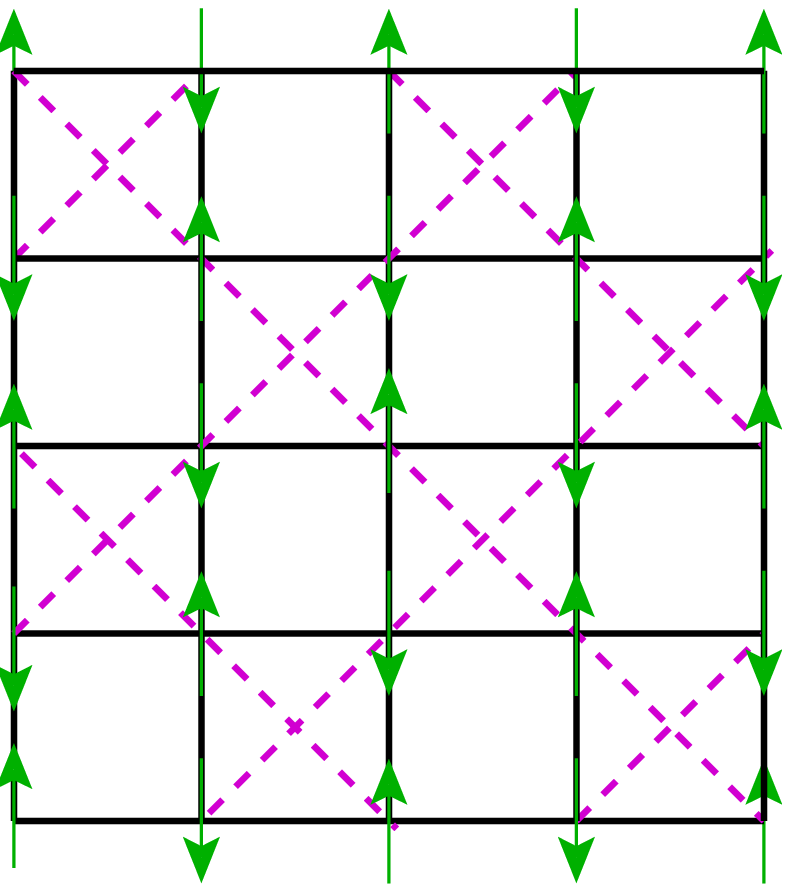}}} \quad 
\subfigure[N\'{e}el$^{\ast}$]{\scalebox{0.4}{\includegraphics{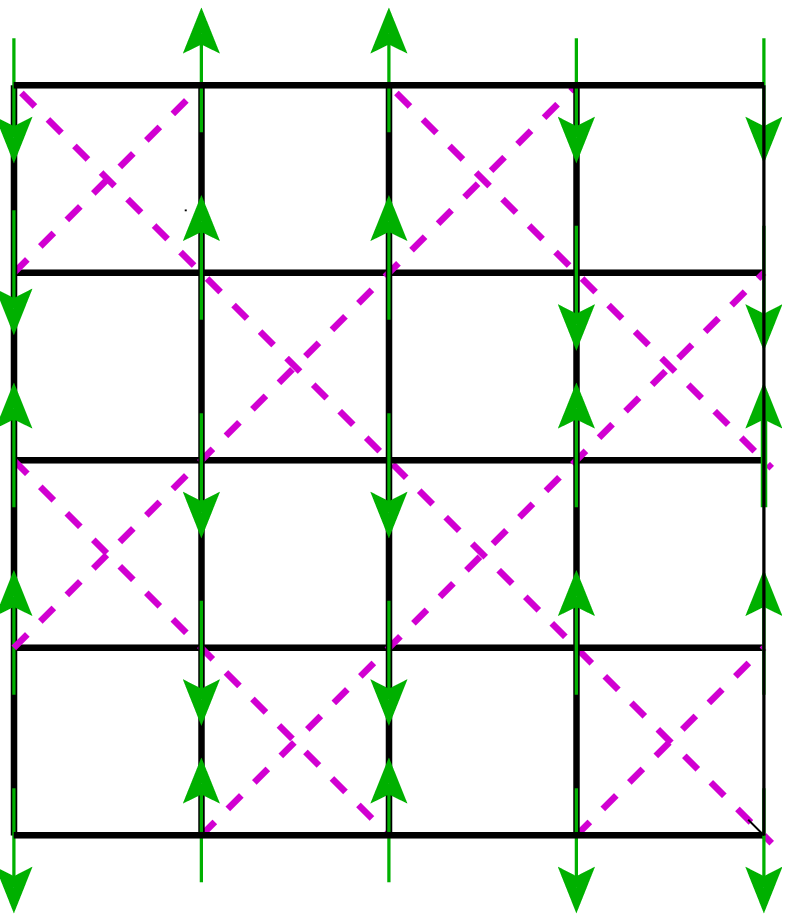}}} 
}
\end{center}
\caption{The $J_{1}$--$J_{2}$ checkerboard model, showing 
  (a) the N\'{e}el state, and (b) one of the two N\'{e}el$^{\ast}$ states.
  The NN $J_{1}$ bonds are shown as solid (black) lines and the NNN (i.e., also NN)
  $J_{2}$ bonds are shown as dashed (magenta) lines.  The arrows
  represent the orientations of the spins on each lattice site 
  for each of the three states shown.}
\label{model_bonds}
\end{figure}

The Hamiltonian of the $J_{1}$--$J_{2}$ model on the checkerboard
lattice is given by
\begin{equation}
H = J_{1}\sum_{\langle i,j \rangle} \mathbf{s}_{i}\cdot\mathbf{s}_{j} + J_{2}\sum_{\langle\langle i,k \rangle\rangle'} 
\mathbf{s}_{i}\cdot\mathbf{s}_{k}\,, \label{H}
\end{equation}
where the indices runs over all sites of a 2D square lattice, such
that the sum over $\langle i,j \rangle$ counts every NN pair (once and once only)
and the sum over $\langle\langle i,k\rangle\rangle'$ counts each NNN pair in the
checkerboard pattern (once and once only) shown in
Fig.~\ref{model_bonds}, such that alternate basic square plaquettes
have either two diagonal bonds or none.  Each site $i$ of the lattice
now carries a particle with spin $s=1$ described by a spin operator ${\bf
  s}_{i}=(s_{i}^{x},s_{i}^{y},s_{i}^{z})$.

The lattice and exchange bonds of the anisotropic planar pyrochlore
model are thus shown in Fig.~\ref{model_bonds}, from which one sees
clearly how it may alternatively be construed as comprising crossed
(diagonal) sets of chains, along which the intrachain exchange coupling
constant is $J_{2}$, coupled by both vertical and horizontal
interchain exchange bonds of strength $J_{1}$.  Both bonds here are
assumed to be AFM in nature (i.e., $J_{1}>0$ and $J_{2}>0$) and hence
to act to frustrate one another.  Clearly, the model interpolates
smoothly between the HAFM on the square lattice (when $\kappa \equiv
J_{2}/J_{1} =0$) and decoupled 1D isotropic HAFM chains (when $\kappa
\rightarrow \infty$).  When $\kappa=1$, in between these limiting
cases, we have the isotropic HAFM on the checkerboard lattice, alias
the isotropic planar pyrochlore.  With no loss of generality,
henceforth we choose $J_{1} \equiv 1$ to set the overall energy scale.

Classically (i.e., when $s \rightarrow \infty$) the model has a single
($T=0$) GS phase transition at $\kappa = \kappa_{{\rm cl}} = 1$.  For
$\kappa < 1$ the GS phase is the N\'{e}el state illustrated in
Fig.~\ref{model_bonds}(a), for which the ordering along the diagonal
chains is hence ferromagnetic (FM) in nature.  The GS energy per spin
of this classical N\'{e}el state is thus $E^{{\rm cl}}/N =
s^{2}(-2J_{1}+J_{2}$).  For $\kappa > 1$ the corresponding GS phase is
now infinitely degenerate.  It comprises a set of collinear states in
which every checkerboard diagonal chain of sites connected by $J_{2}$
bonds has N\'{e}el AFM ordering, but where each chain can slide along
its own length without changing the overall energy.  In other words, the
infinite degeneracy in this set is such that the spins along any one
row or column can be arbitrarily assigned.  All such states then have
the same classical energy per spin, $E^{{\rm cl}}/N = -s^{2}J_{2}$,
independent of the value of $J_{1}$.  Among this infinite family of
states is the so-called N\'{e}el$^{\ast}$ state shown in
Fig.~\ref{model_bonds}(b), which has pairwise or doubled AFM ordering
of the type $\cdots \uparrow \uparrow \downarrow \downarrow \uparrow
\uparrow \downarrow \downarrow \cdots$, along every row and column.
Thus, the single spin $\uparrow$ or $\downarrow$ of the N\'{e}el state
is replaced by the two-site unit $\uparrow\uparrow$ or
$\downarrow\downarrow$ in the N\'{e}el$^{\ast}$ state.  The
N\'{e}el$^{\ast}$ state retains a double degeneracy.  This can easily
be seen, for example, from Fig.~\ref{model_bonds}(b), where the
N\'{e}el$^{\ast}$ state exhibits two forms of empty plaquettes
(namely, those with four parallel spins and those with two pairs of
antiparallel spins), the roles of which can be interchanged.

In our previous \cite{Bishop:2012_checkerboard} CCM analysis of the
$s=\frac{1}{2}$ version of the model we found a GS phase diagram with
marked differences to its classical ($s \rightarrow \infty$)
counterpart.  Although the quasiclassical state with N\'{e}el AFM
ordering remains the GS phase for low enough values of $\kappa$ (viz.,
$\kappa < \kappa_{c}^{1} \approx 0.80 \pm 0.01$), we found that {\it
  none} of the infinitely degenerate set of AFM states in the
classical model (which form the GS phase in that case for $\kappa >
1$) can survive the quantum fluctuations present in the
$s=\frac{1}{2}$ case to form a stable GS phase.  We found instead two
different forms of valence-bond crystalline (VBC) ordering for $\kappa
> \kappa_{c}^{1}$.  Firstly, in the region $\kappa_{c}^{1} < \kappa <
\kappa_{c}^{2} \approx 1.22 \pm 0.02$ we found the stable GS phase to
exhibit plaquette VBC (PVBC) order, while for {\it all} $\kappa > \kappa_{c}^{2}$ the ordering changes to a crossed-dimer VBC (CDVBC)
variety.  We found that both transitions are probably direct ones,
although, as usual, we could not entirely rule out very narrow
coexistence regions confined, respectively, to $0.79 \lesssim \kappa
\lesssim 0.81$ and $1.20 \lesssim \kappa \lesssim 1.22$.

In view of our earlier discussion in Sec.~\ref{intro}, it is now of
great interest to perform a comparable analysis of the $s=1$ version
of the model, the results of which we present in Sec.~\ref{results}
after a brief discussion of the CCM formalism used to obtain them.

\section{The coupled cluster method}
\label{ccm_section}
Since the CCM is well documented elsewhere (see, e.g.,
Refs.~\cite{Bi:1991_TheorChimActa,Bishop:1998,Ze:1998,Fa:2004}) we give only a brief outline
here.  We note that it is particularly well suited for the study of such
highly frustrated magnets as we consider here, for which
alternative methods, such as quantum Monte Carlo (QMC) or exact
diagonalization (ED) techniques, run into severe problems.  Thus, QMC
methods suffer in such cases from the infamous ``minus-sign problem,''
while ED methods are often restricted to too small lattices to be able
to sample with sufficient accuracy the details of the often very
subtle ordering that is present, even when state-of-the-art calculations
are performed with the largest computational resources available.
While both QMC and ED calculations are performed on lattices with a
finite number $N$ of spins, and hence require finite-size scaling to
obtain the $N \rightarrow \infty$ limit required, the CCM, as we
described below, is a size-extensive method that automatically works
in the (infinite-lattice) thermodynamic limit from the outset, at every
level of approximation.  Since such approximations, as we will see
below, can be defined in rigorous hierarchical schemes, the only final
extrapolation needed is to the (exact) limit in any such scheme.
Furthermore, at the highest levels of approximation feasible with
available computational resources, results for physical quantities are
often already well converged, as our specific results in
Sec.~\ref{results} will show.

The CCM starts with the choice of a suitable model state (or reference
state), $|\Phi \rangle$, on top of which the quantum correlations
present in the exact GS phase under study can be systematically
incorporated later, as we described below.  For the present model
obvious choices are the N\'{e}el and N\'{e}el$^{\ast}$ states shown in
Fig.~\ref{model_bonds}.  At the quasiclassical level we expect these
might prove good candidate CCM model states for the regions $\kappa
\lesssim 1$ and $\kappa \gtrsim 1$, respectively.  Of course, for the
latter regime there is an infinite family of classically degenerate
states from which to choose.  We note, in this context, however, that
at the $O(1/s)$ level in a quasiclassical expansion in powers of
$1/s$, a fourfold set of states is selected \cite{Tchernyshyov:2003}
by the {\it order by disorder} mechanism
\cite{Villain:1977,Shender:1982} to lie lowest in energy.  These
include the (doubly degenerate) N\'{e}el$^{\ast}$ states as well as
the two (row and column) striped AFM states, in which alternating rows
or columns have spins aligned $\uparrow$ or $\downarrow$.

Once a model state $|\Phi \rangle$ is chosen, the exact GS ket- and
bra-state wave functions that satisfy the corresponding
Schr\"{o}dinger equations,
\begin{equation}
H|\Psi\rangle = E|\Psi\rangle\,; \quad \langle \tilde{\Psi}|H = E\langle \tilde{\Psi}|\,, \quad
\end{equation}
are parametrized as 
\begin{equation}
|\Psi \rangle = {\rm e}^{S} |\Phi \rangle\,; \quad \langle \tilde{\Psi}| = \langle\Phi|\tilde{S}{\rm e}^{-S}\,, \label{para_ket_bra_eqs}
\end{equation}
where we use the intermediate normalization scheme for $|\Psi\rangle$,
such that $\langle\Phi|\Psi\rangle = \langle\Phi|\Phi\rangle \equiv 1$, and
then for $\langle \tilde{\Psi}|$ choose its normalization such that
$\langle\tilde{\Psi}|\Psi\rangle = 1$.  The correlation operators
$S$ and $\tilde{S}$ are decomposed in terms of exact sets of
multiparticle, multiconfigurational creation and destruction operators,
$C^{+}_{I}$ and $C^{-}_{I} \equiv
(C^{+}_{I})^{\dagger}$, respectively, as
\begin{equation}
S=\sum_{I \neq 0}{\cal S}_{I}C^{+}_{I}\,; \quad \tilde{S} = 1 + \sum_{I \neq 0}\tilde{{\cal S}}_{I}C^{-}_{I}\,,   \label{corr_operators}
\end{equation}
where $C^{+}_{0} \equiv 1$, the identity operator, and $I$ is a set
index describing a complete set of single-particle configurations for
all of the particles.  The reference state $|\Phi \rangle$ thus acts as
a fiducial (or cyclic) vector, or generalized vacuum state, with respect to
the complete set of creation operators $\{C^{+}_{I}\}$, which are hence required to
satisfy the conditions $\langle \Phi|C^{+}_{I} = 0 =
C^{-}_{I}|\Phi\rangle\,, \forall I \neq 0$.

In order to consider each site on the spin lattice to be equivalent to
all others, whatever the choice of state $|\Phi \rangle$, it is
convenient to form a passive rotation of each spin so that in its own
local spin-coordinate frame it points in the downward, (i.e., negative
$z$) direction.  Clearly, such choices of local spin-coordinate frames
leave the basic SU(2) spin commutation relations unchanged, but have
the nice effect that the $C^{+}_{I}$ operators can be expressed as
products of single-spin raising operators $s^{+}_{k} \equiv s^{x}_{k}
+ is^{y}_{k}$, such that $C^{+}_{I} \equiv s^{+}_{k_{1}}s^{+}_{k_{2}}\cdots
s^{+}_{k_{n}};\;n=1,2,\cdots,2sN$.

The complete set of multiparticle correlation coefficients
$\{{\cal S}_{I},{\tilde{\cal S}}_{I}\}$ may now be evaluated by extremizing the
energy expectation value
$\bar{H}\equiv\langle\tilde{\Psi}|H|\Psi\rangle=\langle\Phi|{\tilde{S}}{\rm e}^{-S}H{\rm e}^{S}|\Phi\rangle$, with respect to each of
them, $\forall I \neq 0$.  Variation with respect to each coefficient
${\tilde{\cal S}}_{I}$ yields the coupled set of nonlinear equations,
\begin{equation}
\langle\Phi|C^{-}_{I}{\rm e}^{-S}H{\rm e}^{S}|\Phi\rangle=0\,, \quad \forall I \neq 0\,,  \label{nonlinear_eq}
\end{equation}
for the coefficients $\{{\cal S}_{I}\}$, while variation with respect to
each coefficient ${\cal S}_{I}$ yields the corresponding set of linear
equations,
\begin{equation}
\langle\Phi|\tilde{S}({\rm e}^{-S}H{\rm e}^{S} -
E)C^{+}_{I}|\Phi\rangle=0\,, \quad \forall I \neq 0\,,   \label{ket_linearEqs}
\end{equation}
for the coefficients $\{{\tilde{\cal S}}_{I}\}$, once the coefficients $\{{\cal S}_{I}\}$ have been calculated from Eq.~(\ref{nonlinear_eq}), and where in
Eq.~(\ref{ket_linearEqs}) we have used Eqs.~(\ref{para_ket_bra_eqs})
and (\ref{corr_operators}) to introduce the GS energy $E$.

Up till now everything has been exact.  In practice, of course,
approximations need to be made, and these are made within the CCM by
restricting the set of indices $\{I\}$ retained in the expansions of
Eq.~(\ref{corr_operators}) for the otherwise exact correlation
operators $S$ and $\tilde{S}$.  Some specific such hierarchical scheme
are described below.  It is important to realize, however, that no
further approximations are made.  In particular, the method is
guaranteed by the use of the exponential parametrizations in
Eq.~(\ref{para_ket_bra_eqs}) to be size-extensive at every level of
truncation, and hence we work from the outset in the $N \rightarrow
\infty$ limit.  Similarly, the important Hellmann-Feynman theorem is
similarly exactly obeyed at every level of truncation.  Lastly, when
the similarity-transformed Hamiltonian ${\rm e}^{-S}H{\rm e}^{S}$ in
Eqs.~(\ref{nonlinear_eq}) and (\ref{ket_linearEqs}) is expanded in
powers of $S$ using the well-known nested commutator expansion, the
fact that $S$ contains only spin-raising operators guarantees that the
otherwise infinite expansion actually terminates at a finite order, so
that no further approximations are needed.

Once an approximation has been chosen and the retained coefficients
$\{{\cal S}_{I},{\tilde{\cal S}}_{I}\}$ calculated from
Eqs.~(\ref{nonlinear_eq}) and (\ref{ket_linearEqs}), any GS quantity
can, in principle, be calculated.  For example, the GS energy $E$ can
be calculated in terms of the coefficients $\{{\cal S}_{I}\}$ alone, as
$E=\langle\Phi|{\rm e}^{-S}H{\rm e}^{S}|\Phi\rangle$, while the average
on-site GS magnetization (or magnetic order parameter) $M$ needs both
sets $\{{\cal S}_{I}\}$ and $\{{\tilde{\cal S}}_{I}\}$ for its evaluation as
$M = -\frac{1}{N}\langle\Phi|\tilde{S}{\rm
  e}^{-S}\sum^{N}_{k=1}s^{z}_{k}{\rm e}^{S}|\Phi\rangle$, in terms of
the rotated local spin-coordinate frames defined above.

In our previous work for the $s=\frac{1}{2}$ model
\cite{Bishop:2012_checkerboard} we employed the well-known and
well-tested localized LSUB$m$ CCM approximation scheme (see
Refs.~\cite{Ze:1998,Fa:2004}).  At the $m$th level of approximation it
includes all spin clusters described by multispin configurations in
the index set $\{I\}$ that may be defined over any possible lattice
animal (or polyomino) of size $m$ on the lattice.  Such a lattice
animal is defined in the usual graph-theoretic sense to be a
configured set of contiguous sites on the lattice, in which every site
in the configuration is adjacent (in the NN sense) to at least
one other site.  Clearly, as $m \rightarrow \infty$ the LSUB$m$
approximation becomes exact.  The definition of contiguity employed
above depends itself on the choice of ``geometry'' of the lattice,
i.e., on the definition of what is meant by a NN pair.  Just as in our
previous treatment \cite{Bishop:2012_checkerboard} of the
$s=\frac{1}{2}$ version of the present $s=1$ model, we assume the
fundamental checkerboard geometry to define the retained
configurations, in which pairs of sites connected either by a $J_{1}$
bond or by a $J_{2}$ bond are defined to be contiguous (or as NN pairs
for the sake of defining a lattice animal of a given size).  Although
the number of retained configurations at a given $m$th level of
approximation is larger in the checkerboard geometry than in the
corresponding square-lattice geometry (for which pairs connected by
$J_{2}$ bonds would be NNN pairs), the advantage is that the former
choice retains many of the symmetries of the checkerboard-lattice
model at all levels of approximation that would be lost in the latter
choice.

At a given $m$th level of LSUB$m$ approximation (with {\it any} fixed
choice of underlying geometry to define contiguity) the number,
$N_{f}$, of such distinct (i.e., under the symmetries of the lattice
and specified model state) fundamental spin configurations is lowest
for $s=\frac{1}{2}$ and rises steeply as $s$ increases.  This is
because each downward-pointing (in the rotated local frame) spin on
each site $k$ may be operated upon by the spin-raising operator
$s^{+}_{k}$ up to 2$s$ times.  Thus each site index $k_{i}$ in the
operators $C^{+}_{I} \equiv s^{+}_{k_{1}}s^{+}_{k_{2}}\cdots
s^{+}_{k_{n}}$ may be repeated up to a maximum of 2$s$ times.  For
such $s>\frac{1}{2}$ cases, where individual indices may be repeated,
an alternative, so-called SUB$n$--$m$, CCM scheme has been used.  This
scheme doubly restricts the configured clusters included to contain no
more than $n$ spin-flips (where each spin-flip requires the action of
an $s^{+}_{k}$ operator acting once) spanning a range of no more than
$m$ contiguous sites on the lattice.  We then set $m=n$ and employ
here the SUB$n$--$n$ scheme.  Clearly, the LSUB$m$ scheme is
equivalent to the SUB$n$--$m$ scheme when $n=2sm$ for particles of
spin $s$.  For the $s=\frac{1}{2}$ case only, LSUB$m$ $=$ SUB$m$--$m$,
whereas for the $s=1$ case LSUB$m \equiv$ SUB2$m$--$m$.  We note that
the corresponding numbers, $N_{f}$, of fundamental configurations at a
given SUB$n$--$n$ level are higher for the $s=1$ case than for the
$s=\frac{1}{2}$ case.  Thus, whereas for the $s=\frac{1}{2}$ case we
were able to perform LSUB$m$ calculations with $m \leq 10$ previously
\cite{Bishop:2012_checkerboard}, with similar supercomputer
resources available we are now only able to perform SUB$n$--$n$
calculations for the $s=1$ case that are restricted to $n \leq 8$.  As
before \cite{Bishop:2012_checkerboard} we similarly use massively
parallel computation \cite{ccm} to derive and solve the
corresponding coupled sets of CCM equations (\ref{nonlinear_eq}) and
(\ref{ket_linearEqs}).

As a last step we need to extrapolate the approximate SUB$n$--$n$
results thus obtained to the exact $n \rightarrow \infty$ limit.  Just
as for the $s=\frac{1}{2}$ version of the model
\cite{Bishop:2012_checkerboard} we use for the $s=1$ version the very
well tested and very robust approximation scheme,
\begin{equation}
\frac{E(n)}{N} = a_{0}+a_{1}n^{-2}+a_{2}n^{-4}\,,     \label{extrapo_E}
\end{equation}
for the GS energy per
spin~\cite{Schm:2006,Darradi:2008,Reuther:2011_J1J2J3mod,Bishop:2000,Kr:2000,Fa:2001_PRB64,Darradi:2005_PRB72,Farnell:2008,Bi:2008_PRB_J1xxzJ2xxz,Bi:2008_J1J1primeJ2,Bi:2008_EPL,Bi:2008_JPCM_J1xxzJ2xxz_s1,Bi:2009_SqTriangle,Richter:2010:J1J2mod_FM,Bishop:2010_UJack,Bishop:2010_KagomeSq,Bishop:2010_UJack_GrtSpins,DJJF:2011_honeycomb,Gotze:2011_kagome,PHYLi:2011_SqTriangle_grtSpins,PHYLi:2012_honeycomb_J1neg,Bishop:2012_honeyJ1-J2,Bishop:2012_Honeycomb_J2neg,Bishop:2012_checkerboard,Li:2012_honey_full,Li:2012_anisotropic_kagomeSq,RFB:2013_hcomb_SDVBC,Li:2013_chevron}.
Not surprisingly, the GS expectation values of other physical
observables both tend to converge more slowly than the GS energy, and
with leading exponents that can also depend on the model and regime
under study.  More specifically, the amount of frustration present can
then often determine the scaling.

For example, for most systems with no or moderate amounts of
frustration present, the magnetic order parameter $M$ has been widely
found
\cite{Kr:2000,Bishop:2000,Fa:2001_PRB64,Darradi:2005_PRB72,Bi:2009_SqTriangle,Bishop:2010_UJack,Bishop:2010_KagomeSq,Bishop:2010_UJack_GrtSpins}
to obey a scaling law with leading power $1/n$ (rather than with
$1/n^{2}$ as for the GS energy).  In such cases an extrapolation scheme
of the form
\begin{equation}
M(n) = b_{0}+b_{1}n^{-1}+b_{2}n^{-2}\,,   \label{M_extrapo_standard}
\end{equation}
works well.  However, for systems which are close to a quantum critical point (QCP) or for
which the magnetic order parameter for the phase being studied is zero
or close to zero, the extrapolation scheme of
Eq.~(\ref{M_extrapo_standard}) tends always to overestimate the
(extrapolated value of the) order parameter and hence also to
predict a somewhat too large value for the critical strength of the
frustrating interaction that drives the transition under study.  In
such cases much evidence has by now been accumulated that a scaling law
with leading power $1/n^{1/2}$ works much better to fit the
SUB$n$--$n$ data.  In such cases we then use the alternative
well-studied extrapolation scheme
~\cite{Schm:2006,Darradi:2008,Reuther:2011_J1J2J3mod,Bi:2008_PRB_J1xxzJ2xxz,Bi:2008_J1J1primeJ2,Bi:2008_EPL,Bi:2008_JPCM_J1xxzJ2xxz_s1,Richter:2010:J1J2mod_FM,DJJF:2011_honeycomb,Gotze:2011_kagome,PHYLi:2011_SqTriangle_grtSpins,PHYLi:2012_honeycomb_J1neg,Bishop:2012_honeyJ1-J2,Bishop:2012_Honeycomb_J2neg,Bishop:2012_checkerboard,Li:2012_honey_full,Li:2012_anisotropic_kagomeSq,RFB:2013_hcomb_SDVBC,Li:2013_chevron},
\begin{equation}
M(n) = c_{0}+c_{1}n^{-1/2}+c_{2}n^{-3/2}\,.    
\label{M_extrapo_frustrated}
\end{equation} 

For any physical observable $P$ of any spin-lattice model being
studied by the CCM, we may obviously always test for the correct
scaling by first fitting the SUB$n$--$n$ results to a form
\begin{equation}
P(n) = p_{0}+p_{1}n^{-\nu}\, ,    
\label{M_extrapo_nu}
\end{equation} 
where the leading exponent $\nu$ is also a fitting parameter.  For the
GS energy $E$ such fits generally yield a fitted value of $\nu$ very
close to 2 for a wide variety of both unfrustrated and (even highly)
frustrated systems in different phases, as is also the case here.
Such a preliminary fit then justifies the use of Eq.~(\ref{extrapo_E})
to find the extrapolated value of $E/N$.  A similar preliminary
analysis for the order parameter $M$ can be done in specific cases to
justify the use of either Eq.~(\ref{M_extrapo_standard}) or
Eq.~(\ref{M_extrapo_frustrated}) to find the extrapolated value of
$M$.

For the model at hand we have performed extrapolations for all of the
physical observables calculated using each of the SUB$n$--$n$ data
sets with $n=\{2,4,6,8\}$ and $n=\{4,6,8\}$ (and also $n=\{2,4,6\}$,
even though this set is clearly not a preferred set), as a further
test of the robustness of the schemes used.  In each case we find very
similar results for the extrapolated values, thereby lending credence
to the extrapolation schemes used to find them.

\section{Results}
\label{results}
\begin{figure}
\begin{center}
  \includegraphics[width=6.5cm,angle=270]{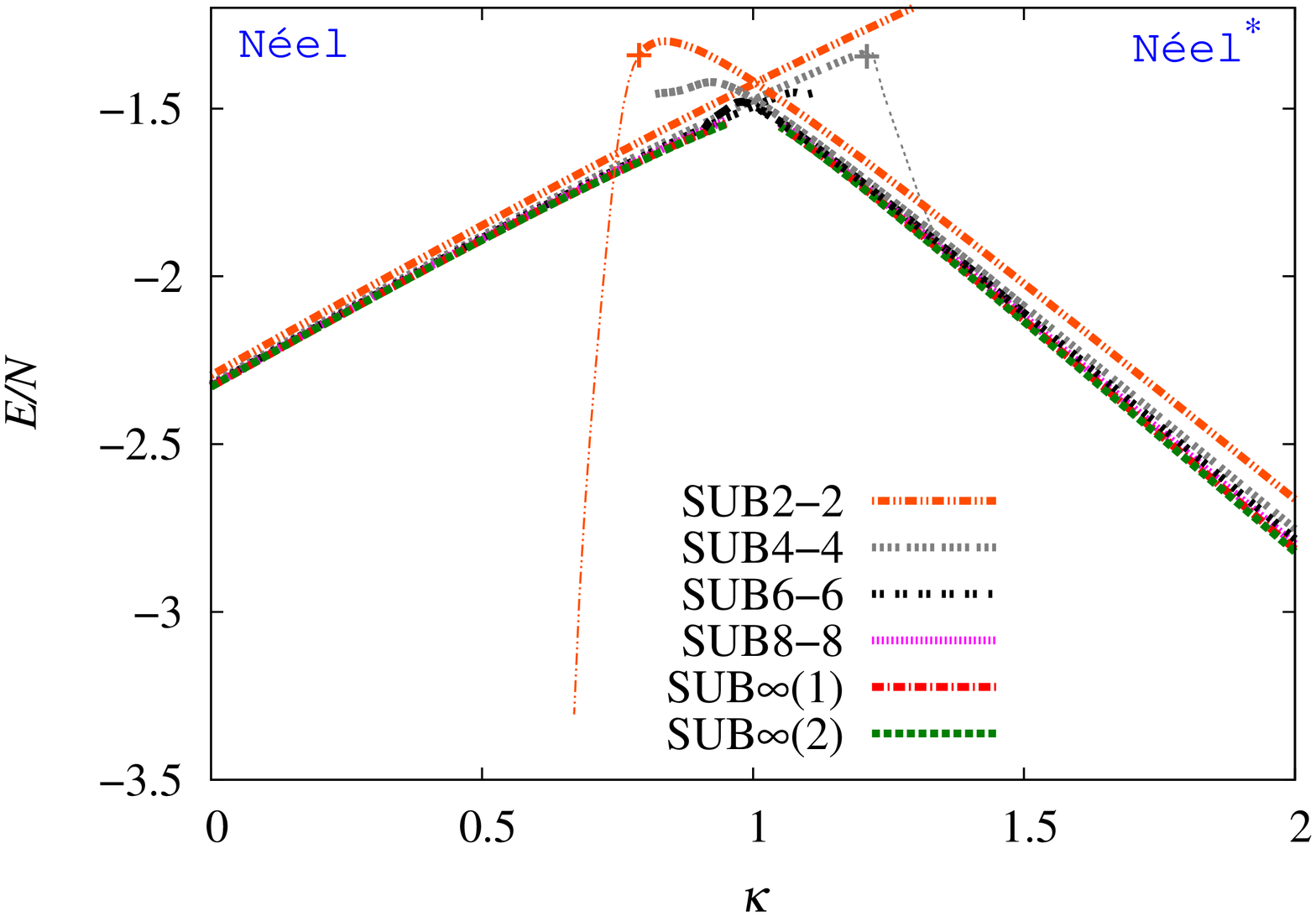}
  \caption{CCM results for the GS energy per spin,
    $E/N$, as a function of the frustration parameter, $\kappa \equiv
    J_{2}/J_{1}$, for the spin-1 $J_{1}$--$J_{2}$ Heisenberg
    antiferromagnet (with $J_{1} \equiv 1$) on the checkerboard
    lattice.  The results for the SUB$n$--$n$ approximations with
    $n=\{2,4,6,8\}$ based on the N\'{e}el state (left curves) and
    N\'{e}el$^{\ast}$ state (right curves) as the respective CCM model
    state are shown, together with the corresponding SUB$\infty(k)$
    extrapolations obtained from using Eq.~(\ref{extrapo_E}) together
    with the respective SUB$n$--$n$ data sets: $k = 1$,
    $n=\{2,4,6,8\}$; and $k=2$, $n=\{4,6,8\}$.
    All SUB$n$--$n$ solutions are shown out to their respective
    (approximately determined) mathematical termination points; and
    the plus ($+$) symbols mark those points where the corresponding
    solutions have vanishing order parameter $M=0$ (and see Fig.~\ref{M_s1}(a)).  Those portions of the curves beyond the plus
    ($+$) symbols, shown with thinner curves, indicate unphysical
    regions, where $M<0$ for these approximations (and see text for
    further details).}
\label{E_s1}
\end{center}
\end{figure}

We now present our CCM results for the spin-1 $J_{1}$--$J_{2}$ model
on the checkerboard lattice, using both the N\'{e}el and
N\'{e}el$^{\ast}$ states shown in Fig.\ \ref{model_bonds} as
model states, and employing the SUB$n$--$n$ truncation scheme with $n
\leq 8$.  Figure \ref{E_s1} firstly shows the results for the GS energy per
spin, $E/N$.  We display both ``raw'' SUB$n$--$n$ results and
extrapolations to the $n \rightarrow \infty$ limit based on the use of
Eq.\ (\ref{extrapo_E}) with either the data set $n=\{2,4,6,8\}$ or
$n=\{4,6,8\}$.
The first thing to notice is how well converged the results are for
both sets of results based on the N\'{e}el state (left curves) and
N\'{e}el$^{\ast}$ state (right curves).  Secondly, we note that,
exactly as for the $s=\frac{1}{2}$ case studied previously
\cite{Bishop:2012_checkerboard}, both sets of curves display
termination points, at particular values of $\kappa$, namely, an upper
one for the N\'{e}el curves and a lower one for the N\'{e}el$^{\ast}$
curves, beyond which no real solutions exist.  The termination points
themselves depend on the particular SUB$n$--$n$ approximation used.
What is generally observed is that as the truncation index $n$ is
increased the range of values of the frustration parameter $\kappa$
over which the corresponding SUB$n$--$n$ approximations have real
solutions decreases, as may clearly be seen from Fig.~\ref{E_s1}.
Such CCM termination points are by now well understood
\cite{Fa:2004,Bishop:2010_UJack}.  Indeed, they provide a clear first
signal of the corresponding QCPs that exist in the system under study.
However, we note that it is computationally expensive to obtain the
actual termination point with great accuracy, since the CCM
SUB$n$--$n$ solutions require increasingly more computing power the
nearer one approaches a termination point.  Thus, particularly for the
higher values of the truncation index $n$, it is almost certain that
real solutions exist for slightly larger ranges of $\kappa$ than those
shown.

In the vicinity of any such mathematical CCM SUB$n$--$n$ (or LSUB$m$)
solution termination points it is commonly found, as is the case here
too as we shall see explicitly below, that the corresponding solutions
themselves become unphysical in the sense that the respective values
of the magnetic order parameter (namely, the local on-site
magnetization, $M$) become negative.  These values where $M$ changes
sign in this way, determined as discussed below, are denoted by plus
sign ($+$) symbols in Fig.\ \ref{E_s1}, and the unphysical regions
beyond them, out to the mathematical termination points, where $M<0$
are shown by thinner curves than the corresponding physical regions
where $M>0$ that are marked by thicker curves.

We note from Fig.\ \ref{E_s1} that the corresponding SUB$n$--$n$
curves for $E/N$ based on the N\'{e}el and the N\'{e}el$^\ast$ states,
intersect one another for the same level $n$ of truncation, at least
for $n \leq 6$.  It is almost certain that this is true too for the $n=8$
curves, although it is computationally expensive to find the
solutions in this region, as indicated above.  Furthermore, we note
that as the truncation index $n$ increases, the angle of intersection
of the corresponding SUB$n$--$n$ N\'{e}el and N\'{e}el$^{\ast}$ curves
decreases.  Indeed, the corresponding extrapolated curves seem highly
likely to join smoothly.  Thus, there are clear preliminary
indications that the first-order phase transition at $\kappa_{{\rm
    cl}}=1$ in the classical ($s \rightarrow \infty$) version of the
model might even become a continuous second-order transition in its
$s=1$ counterpart.  The energy crossing point of the N\'{e}el and
N\'{e}el$^{\ast}$ results for the $s=1$ model is estimated to be
$\kappa_{c_{1}} \approx 1.00 \pm 0.01$ based on our CCM SUB$n$--$n$
results for $n=\{2,4,6\}$ and a corresponding extrapolation based on
these results.

Thus, based on the energy results alone, the $s=1$ model seems to
share this transition point at which N\'{e}el order melts with its
classical ($s \rightarrow \infty$) counterpart.  Nevertheless, there
are first indictions that the classical first-order transition at
$\kappa_{{\rm cl}}=1$ might become of second-order type for the $s=1$
model.  By contrast, as was shown earlier
\cite{Bishop:2012_checkerboard}, in the $s=\frac{1}{2}$ model N\'{e}el
order melts at a lower QCP, namely $\kappa_{c}^{1} \approx 0.80 \pm
0.01$, at which point it gives way not to the quasiclassical
N\'{e}el$^{\ast}$ state (or any other of the infinitely degenerate family of
classical states) but to a PVBC-ordered state.

In order to investigate the overall level of accuracy of our results
it is worthwhile to examine the two special limiting cases of
$\kappa=0$ (square-lattice HAFM) and $\kappa \rightarrow \infty$
(decoupled 1D HAFM chains).  Thus, firstly, based on the N\'{e}el
state as the CCM model state, our extrapolated results for the GS
energy per spin for the spin-1 square-lattice HAFM are $E(\kappa=0)/N
\approx -2.3287$ based on the SUB$n$--$n$ results using the data set
$n=\{4,6,8\}$ and $E(\kappa=0)/N \approx -2.3292$ using the corresponding data
set $n=\{2,4,6,8\}$.  These are in
very good agreement with the alternative results $E(\kappa=0)/N \approx -2.3282$
based on third-order spin-wave theory (SWT-3) \cite{Ha:1992} and
$E(\kappa=0)/N=-2.3279(2)$ based on a linked-cluster series expansion
(SE) technique \cite{Zh:1991}.  Our results are also in essentially
exact agreement for this $\kappa=0$ case with previous CCM estimates
that are similar limiting cases of spin-1 $J_{1}$--$J_{2}$ models on
the Union Jack lattice \cite{Bishop:2010_UJack_GrtSpins} and an
anisotropic triangular lattice \cite{PHYLi:2011_SqTriangle_grtSpins},
both of which reduce to the square-lattice HAFM when $J_{2}=0$.
Secondly, based on the N\'{e}el$^{\ast}$ state as CCM model state our
extrapolated results for the GS energy per spin for the spin-1 1D HAFM
chain are $E(\kappa\rightarrow\infty)/N=-1.3954\kappa$ based on the SUB$n$--$n$ results using the data set
$n=\{4,6,8\}$ and $E(\kappa\rightarrow\infty)/N=-1.3917\kappa$ using
the corresponding set $n=\{2,4,6,8\}$.  In this special 1D limiting
case essentially exact results are available from density-matrix
renormalization group (DMRG) calculations \cite{White:1993_s1_chain},
which yield $E(\kappa \rightarrow \infty)/N=-1.4015\kappa$.  Once
again, in this particularly challenging limit for the current model,
our CCM results are in very good agreement with the DMRG result.

\begin{figure*}
\mbox{
\subfigure[]{\scalebox{0.3}{\includegraphics[angle=270]{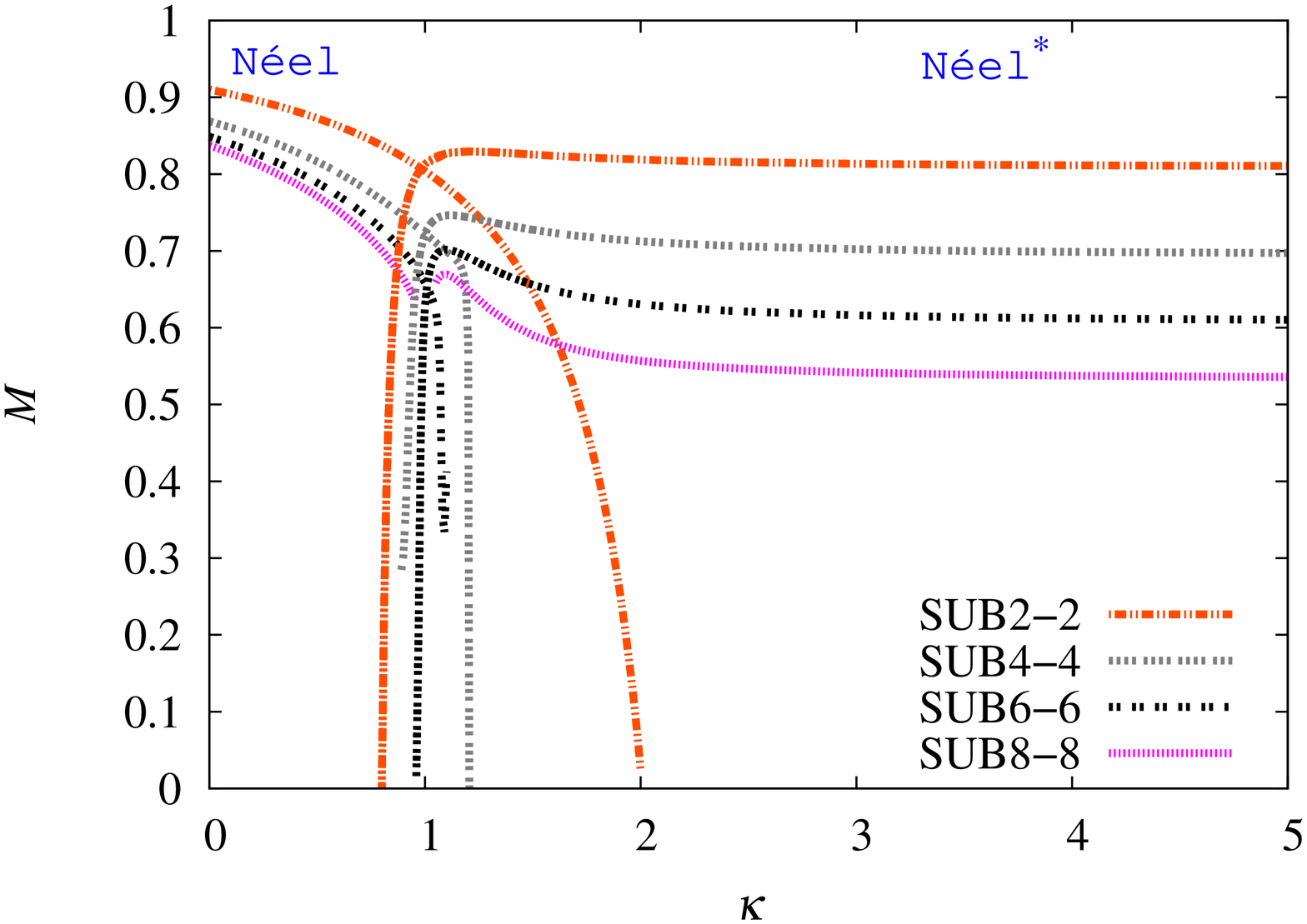}}} \quad 
\subfigure[]{\scalebox{0.3}{\includegraphics[angle=270]{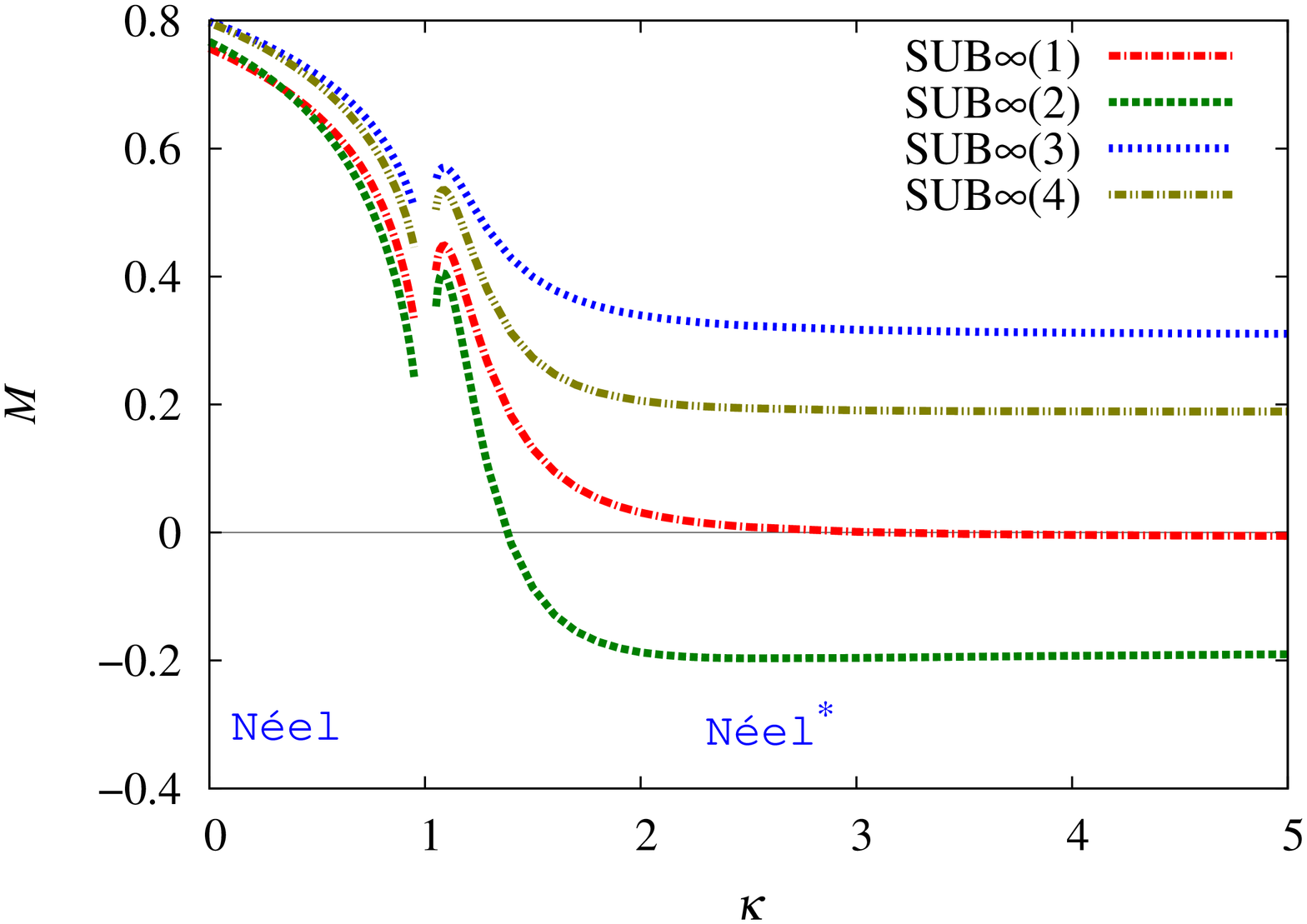}}} 
}
\caption{CCM results for the GS magnetic order parameter, $M$, as a
  function of the frustration parameter, $\kappa \equiv J_{2}/J_{1}$,
  for the spin-1 $J_{1}$--$J_{2}$ Heisenberg antiferromagnet (with
  $J_{1} > 0$) on the checkerboard lattice.  (a) The results for
  the SUB$n$--$n$ approximations with $n=\{2,4,6,8\}$ based on the
  N\'{e}el state (left curves) and N\'{e}el$^{\ast}$ state (right
  curves) as the respective CCM model state.  (b) The corresponding
  SUB$\infty(k)$ extrapolations: those with $k=\{1,2\}$ use the
  extrapolation scheme of Eq.~(\ref{M_extrapo_frustrated}), while
  those with $k=\{3,4\}$ use the extrapolation scheme of Eq.~(\ref{M_extrapo_standard}), in both cases with the respective
  SUB$n$--$n$ data sets: $k = \{1,3\}$, $n=\{2,4,6,8\}$; and $k=\{2,4\}$,
$n=\{4,6,8\}$.}
\label{M_s1}
\end{figure*}

In Fig.~\ref{M_s1} we now show our corresponding results for the GS
magnetic order parameter, $M$, to those shown in Fig.~\ref{E_s1} for
the GS energy per spin, $E/N$.  
Firstly, in Fig.~\ref{M_s1}(a) we show the ``raw'' SUB$n$--$n$ results
based on both the N\'{e}el state (left curves) and N\'{e}el$^{\ast}$
state (right curves) used as the CCM model state.  We note that the
plus sign ($+$) symbols shown in Fig.~\ref{E_s1} for the corresponding
energy results correspond precisely to the respective points in
Fig.~\ref{M_s1}(a) at which $M$ passes through zero.  One sees clearly
that in the vicinity of $\kappa=1$, the SUB$n$--$n$ curves for $M$
become increasingly steep as $n$ increases in value, thereby also
demonstrating why we find increased difficulty in obtaining
solutions in this region, which is also close to the corresponding
termination points discussed previously.  The raw SUB$n$--$n$ results
show very clearly the existence of a phase transition with a QCP at or
very close to $\kappa=1$, just as did the corresponding results for
the GS energy shown in Fig.~\ref{E_s1}.

In Fig.~\ref{M_s1}(b) we also show the extrapolations of the results
in Fig.~\ref{M_s1}(a), based on both the extrapolation schemes of
Eqs.~(\ref{M_extrapo_standard}) and (\ref{M_extrapo_frustrated}), and
using each of the SUB$n$--$n$ data sets $n=\{2,4,6,8\}$ and
$n=\{4,6,8\}$ in the two cases, for completeness and the sake of
comparison.  As we noted earlier, fits to each data set for $M$ for
both model states used, of the form of Eq.~(\ref{M_extrapo_nu}), can
also be performed to estimate the leading exponent and hence to
determine which (if either) of the two forms of
Eqs.~(\ref{M_extrapo_standard}) or (\ref{M_extrapo_frustrated}) is
appropriate in particular cases.  We refer the interested reader to
Ref.~\cite{Bishop:2013_cross-striped}, for example, for a fuller
description of the procedure.  Based on such fits and much prior
experience, the extrapolation scheme of Eq.~(\ref{M_extrapo_standard})
is now clearly indicated for our N\'{e}el-state results over most of
the range of values of $\kappa$ for which solutions exist, except very
near the point $\kappa=1$ where the extrapolation scheme of
Eq.~(\ref{M_extrapo_frustrated}) becomes validated.

In particular, the use of Eq.~(\ref{M_extrapo_standard}) is clearly
indicated at the point $\kappa=0$, corresponding to the special
limiting case of the square-lattice HAFM.  Our corresponding
extrapolated results for the GS magnetic order parameter for the
spin-1 square-lattice HAFM, using Eq.~(\ref{M_extrapo_standard}), are
$M(\kappa=0) \approx 0.796$ based on SUB$n$--$n$ results with $n=\{4,6,8\}$,
and $M(\kappa=0) \approx 0.798$ with $n=\{2,4,6,8\}$.  Once again,
these are in very good agreement with such independent results as
$M(\kappa = 0) \approx 0.804$ based on SWT-3 calculations \cite{Ha:1992} and $M(\kappa =
0) = 0.8039(4)$ from linked-cluster SE calculations \cite{Zh:1991}.
Our current results are also in complete accord with previous CCM estimates
\cite{Bishop:2010_UJack_GrtSpins,PHYLi:2011_SqTriangle_grtSpins}
discussed above in connection with the corresponding results for the GS
energy.

We note from Fig.~\ref{M_s1}(a) that the SUB$n$--$n$ results based on
the N\'{e}el state as CCM model state converge (with increasing values
of $n$) much faster than those for the N\'{e}el$^{\ast}$ state, which
is fully consistent with a smaller value of the leading exponent $\nu$
in the scaling law for the latter results than for the former, over
most of their respective ranges of existence.  In this case, as
explained before, the extrapolation scheme of
Eq.~(\ref{M_extrapo_frustrated}), which has been validated by much prior
experience, is found to be appropriate here too.  The corresponding
extrapolations based on the two data sets $n=\{2,4,6,8\}$ and
$n=\{4,6,8\}$, shown in Fig.~\ref{M_s1}(b), now differ somewhat from
each other.  There are two reasons that come immediately to mind as
possible explanations for this difference.  The first is that the
inclusion of the results with $n=2$ might bias the results somewhat,
in view of them being rather far from the $n \rightarrow \infty$
limit.  For this reason we usually prefer to exclude data with $n=2$.
Another argument for doing so in general for square lattices is that
since the basic square plaquette is an important structural element of
models on the square lattice, approximations with $n \geq 4$ are
preferred {\it a priori}.  However, since we employ here SUB$n$--$n$
approximations based on the checkerboard geometry rather than the
square-lattice geometry, such an argument loses much of its validity.
By contrast, a second reason for the difference in extrapolations
based on the two data sets shown is that 3-parameter fits based on
only 3 inputs are inherently less stable than those based on 4 or
more inputs, and for that reason we generally argue for fits of the latter
type.  Clearly, the above two arguments are in conflict over whether
or not it is better to include or exclude the SUB2-2 results in our
fits.  From both Figs.~\ref{E_s1} and \ref{M_s1}(b) we observe that
the fits are, nonetheless, remarkably similar in both cases for each of the GS
quantities $E/N$ and $M$, with the single exception of results for $M$
in the region $\kappa \gtrsim 1$.

Despite the above caveat, it is abundantly clear that all of the
results for $M$ in Figs.~\ref{M_s1}(a) and \ref{M_s1}(b) point very
strongly to a quantum phase transition at the same value
$\kappa_{c_{1}} \approx 1.00 \pm 0.01$ as indicated by the GS energy
results in Fig.~\ref{E_s1}.  It is clear too that the transition
there is a very sharp one.  The more likely scenario, based on these
results, is of a continuous (but very steep or sharp) second-order
transition at which $M \rightarrow 0$ on both sides of the transition.
Clearly, however, we cannot exclude a weak first-order transition
either in which $M$ approaches the same small (but nonzero) value on
both sides of the transition or in which $M \rightarrow 0$
continuously on one side (likely the N\'{e}el side) and then
discontinuously jumps to a small (but nonzero) value on the other
(most likely, N\'{e}el$^{\ast}$) side.

We turn finally to the results of $M$ in Fig.~\ref{M_s1}(b) based on
the N\'{e}el$^{\ast}$ state in the region $\kappa > 1$.  As indicated
above the appropriate extrapolations in this region should be based on
Eq.~(\ref{M_extrapo_frustrated}), and hence the relevant curves are
those labelled SUB$\infty$(1) and SUB$\infty$(2).  Conflicting reasons
have already been discussed with regard to which of these two
extrapolations to prefer, and it is hard to make an {\it a priori}
decision on this basis.  Nevertheless, we may also appeal to the known
result that the spin-1 1D Haldane chain has $M=0$.  Since this is
precisely our limiting case $\kappa \rightarrow \infty$, on this basis
the extrapolation curve SUB$\infty$(1) is clearly preferred, since it
gives this result within very small errors.

What is clear from Fig.~\ref{M_s1}(b), however, is that whether we use
the SUB$\infty$(1) or SUB$\infty$(2) result, there is clear evidence
that N\'{e}el$^{\ast}$ order is present only over a range of values of
$\kappa_{c_{1}} < \kappa < \kappa_{c_{2}}$ of the frustration
parameter.  The SUB$\infty(2)$ curve shows $M<0$ in this case for
$\kappa > \kappa_{c_{2}} \approx 1.4$ whereas the SUB$\infty(1)$ curve
shows the more physical result that $M=0$ (within very small errors)
for $\kappa > \kappa_{c_{2}} \approx 2.5$.  The existence of this
finite region of stable N\'{e}el$^{\ast}$ order is quite different
both from the $s=\frac{1}{2}$ version of the model (for which
N\'{e}el$^{\ast}$ order exists nowhere as a stable GS phase) and the
classical ($s \rightarrow \infty$) version (for which it co-exists
with an infinite family of states with AFM ordering along crossed
$J_{2}$-coupled chains as the stable GS phase for {\it all} values
$\kappa > \kappa_{{\rm cl}} = 1$).

In Sec.\ \ref{summary} we summarize our results and compare further
the present $s=1$ model with its $s=\frac{1}{2}$ and classical ($s
\rightarrow \infty$) counterparts.

\section{Discussion and summary}
\label{summary}
In this paper we have investigated the GS properties and GS ($T=0$)
phase structure of the frustrated spin-1 AFM $J_{1}$--$J_{2}$ model (with
$J_{1}>0, J_{2} \equiv \kappa J_{1} > 0$) on the checkerboard lattice.
To do so we have employed the CCM in the hierarchical SUB$n$--$n$
approximation scheme carried out to orders $n \leq 8$.  As CCM model
states we have employed two quasiclassical states, namely the N\'{e}el
state and the N\'{e}el$^{\ast}$ state.  The former is the unique GS
phase for $\kappa < \kappa_{{\rm cl}}=1$ for the classical version of
the model, while the latter is one of an infinitely degenerate family
of classical GS phases for $\kappa > \kappa_{{\rm cl}} = 1$.

We find that for the $s=1$ model the GS phase is an AFM
N\'{e}el-ordered state for $\kappa < \kappa_{c_{1}}$, at which point
the staggered N\'{e}el magnetization vanishes.  Our best estimate for
this lower QCP is $\kappa_{c_{1}} \approx 1.00 \pm 0.01$.  On
the one hand this value seems to concur with the classical value.  On
the other hand it is quite different from the $s=\frac{1}{2}$ value of
$\kappa_{c}^{1} \approx 0.80 \pm 0.01$, found previously
\cite{Bishop:2012_checkerboard}.

In the classical version of the model the transition at
$\kappa_{{\rm cl}}=1$ is a direct first-order one (with a discontinuity
in the slope, $dE$/$d\kappa$, of the GS energy there).  We find for
the $s=1$ model that the corresponding transition at $\kappa_{c_{1}}$
is considerably ``softened,'' with the most likely scenario being that
the transition is now a continuous (second-order) one, although we
cannot on the available evidence rule out a weak first-order one.  All
of our evidence is that, as in the classical model, the transition at
$\kappa_{c_{1}}$ is a direct one to a GS phase with another
quasiclassical form of AFM ordering.  We find zero evidence of any
intermediate phase, and we can positively exclude such a phase except
for a very narrow region around $\kappa_{c_{1}}$.  If any such
intermediate phase does exist (which we doubt on the present
evidence), it can do so only over a tiny region confined to the range
$0.99 \lesssim \kappa \lesssim 1.01$.  Similarly, if present at all,
any region of coexistence of N\'{e}el and N\'{e}el$^{\ast}$ ordering is
restricted by our results to a correspondingly narrow range.

In the classical checkerboard model the N\'{e}el ordering that exists
for $\kappa < \kappa_{{\rm cl}} = 1$ gives way for all $\kappa >
\kappa_{{\rm cl}}$ to an infinitely degenerate family of GS phases
characterized by AFM ordering along the (crossed) $J_{2}$ chains.  In
the quasiclassical limit (where one works to leading order in $1/s$)
it has been shown \cite{Tchernyshyov:2003} that quantum fluctuations
select, by the {\it order by disorder} mechanism
\cite{Villain:1977,Shender:1982}, a fourfold family of collinear
states from among all other classically degenerate states.  These
comprise the stripe-ordered and N\'{e}el$^{\ast}$ states, both of
which are doubly degenerate.  In a previous CCM study of the
$s=\frac{1}{2}$ version of the present checkerboard model, the
N\'{e}el$^{\ast}$ state was found to have lower energy than the
stripe-ordered state, and other authors \cite{Starykh:2005} have also
found evidence in favour of the N\'{e}el$^{\ast}$ state for this
model.  Consequently in the present analysis we have also been
motivated to use the N\'{e}el$^{\ast}$ state as a CCM model state to
investigate the phase structure of the $s=1$ checkerboard model for
$\kappa > \kappa_{c_{1}}$.  Nevertheless, further work should be done
in the future to investigate, for example, whether a quasiclassical
striped state might lie lower in energy than the N\'{e}el$^{\ast}$
state for the $s=1$ model.

We have found clear evidence that a N\'{e}el$^{\ast}$ state with
nonzero values of the order parameter exists for the $s=1$ case for
values of the frustration parameter $\kappa_{c_{1}} < \kappa <
\kappa_{c_{2}} \approx 2.0 \pm 0.5$.  This finding, perhaps the most
striking of this study, differs from both the $s=\frac{1}{2}$ and $s
\rightarrow \infty$ (classical) versions of the model.  Firstly, for
the $s=\frac{1}{2}$ model, the previous CCM study
\cite{Bishop:2012_checkerboard} found that the N\'{e}el$^{\ast}$ state
could not survive quantum fluctuations to form a stable GS phase for
{\it any} values of $\kappa$.  When used as a CCM model state,
although solutions could be found for $\kappa > \kappa_{c}^{1}$, the
calculated (i.e., extrapolated) value of its order parameter $M$ was
found to be zero (or negative) everywhere.  By contrast, for the $s=1$
case we find that when the N\'{e}el$^{\ast}$ state is used as the CCM
model state, the calculated value of $M$ is zero (or negative) only
for $\kappa > \kappa_{c_{2}}$ ($> \kappa_{c_{1}}$).  Secondly, by
contrast with the classical checkerboard model, where stable nonzero
N\'{e}el$^{\ast}$ ordering exists for {\it all} values $\kappa >
\kappa_{{\rm cl}}$, its $s=1$ counterpart exists only over a finite range
of values of $\kappa$.

The question thus remains as to what is the stable GS phase of the
$s=1$ checkerboard model for $\kappa > \kappa_{c_{2}}$.  We note that
for its $s=\frac{1}{2}$ counterpart, the previous CCM study
\cite{Bishop:2012_checkerboard} found a PVBC-ordered phase for
$\kappa_{c}^{1} < \kappa < \kappa_{c}^{2} \approx 1.22 \pm 0.02$,
which then gave way to a CDVBC-ordered phase for all $\kappa >
\kappa_{c}^{2}$.  We have also performed very preliminary calculations
(which we will report on, more fully, elsewhere) for these phases for
the $s=1$ model, using the same CCM technique as reported previously
\cite{Bishop:2012_checkerboard} for the $s=\frac{1}{2}$ case.  Very
interestingly, the evidence so far is that for all values $\kappa >
\kappa_{c_{2}}$ the GS phase of the $s=1$ model seems to take PVBC
ordering.  Again, if this result stands up to further scrutiny, the
$s=1$ model will again show distinct differences to both its
$s=\frac{1}{2}$ and $s \rightarrow \infty$
counterparts.

Finally, we make some brief remarks regarding the order of the various
quantum phase transitions exhibited by the checkerboard model, and in
particular whether specific transitions are allowed to be continuous.
Although a general renormalization group theory approach to continuous
critical phenomena itself places only rather weak constraints on the
existence of a continuous phase transition between any two quantum
phases, the traditional (and conventional) Landau-Ginzburg-Wilson
(LGW) approach \cite{Landau:1999,Wilson:1974} places stricter
criteria.  In a nutshell LGW theory places as a necessary condition on a
continuous transition from a phase $\mathcal{P}_{1}$ to a phase $\mathcal{P}_{2}$ that the
symmetry group of phase $\mathcal{P}_{2}$ is a subgroup of that $\mathcal{P}_{1}$
\cite{Landau:1999}.  Physically, the order parameter of some mode of
phase $\mathcal{P}_{1}$ goes to zero at the transition as the mode becomes soft
and macroscopic condensation into it hence occurs, with a consequent
symmetry reduction.

A detailed analysis and description of the various LGW-allowed
continuous phase transitions for the checkerboard model has been given
elsewhere \cite{Starykh:2005}.  In particular it is shown explicitly
that the symmetry group of the N\'{e}el$^{\ast}$ state is a subgroup of
that of the PVBC state.  Hence, our tentative identification of the
QCP at $\kappa_{c_{2}}$ in the $s=1$ checkerboard model as being
between states with N\'{e}el$^{\ast}$ and PVBC order, would be
LGW-allowed as a continuous transition, and it will be interesting to
examine the order of the transition in more detail.  By contrast, it
is not difficult to show (and see, e.g., Ref.~\cite{Starykh:2005} for
explicit details) that the N\'e{e}l$^{\ast}$ and CDVBC states break
{\it different} symmetries of the checkerboard lattice (and hence of
the Hamiltonian), so that the symmetry group of neither state is a
subgroup of the other.  Hence, any continuous phase transition
between states with N\'{e}el$^{\ast}$ and CDVBC ordering would be
LGW-forbidden.  In such a case the most likely scenarios would be
either a direct first-order transition or one that involves an
intermediate coexistence phase with intermediate ordering and bond
modulation.  Reference \cite{Starykh:2005} describes the possible
properties of such a coexistence phase and the nature of the symmetry
breakings at the two transition points from it into each of the pure phases.

The nature of the QCP at $\kappa_{c}^{2}$ in the $s=\frac{1}{2}$ model
is also more open than that at $\kappa_{c_{2}}$ in the $s=1$
model.  Thus, again, while both the CDVBC and PVBC phases are doubly
degenerate (and can thus be described via Ising order parameters), it
is easy to see that they have distinct symmetries \cite{Starykh:2005}.
Thus, again, neither state has a symmetry group which is a subgroup of
that of the other, and a continuous transition between them is
LGW-forbidden.

Lastly, we note too that the transitions at $\kappa_{c}^{1}$ in the
$s=\frac{1}{2}$ model and at $\kappa_{c_{1}}$ in the $s=1$ model, from
the N\'{e}el phase to, respectively, the PVBC phase (for
$s=\frac{1}{2}$) and the N\'{e}el$^{\ast}$ phase (for $s=1$) are both
also LGW-forbidden as continuous transitions.  Thus, for example, the
N\'{e}el$^{\ast}$ (and PVBC) phases both have rotations of
$\frac{1}{2}\pi$ about the centre of any square plaquette containing
four parallel spins as symmetries, which are not shared with the
N\'{e}el phase.  Hence, if the transition at $\kappa_{c_{1}}$ for the
present $s=1$ model is indeed continuous, as seems possible from our
results, its nature has to be sought outside the conventional LGW
paradigm.  One such possibility, which takes us too far afield to study
further here however, is via the deconfinement scenario
\cite{Sentil:2004}.

In conclusion, the $s=1$ checkerboard model has been shown to exhibit
some very interesting features of its GS ($T=0$) phase diagram that
are qualitatively different to those of both its $s=\frac{1}{2}$ and
$s \rightarrow \infty$ (classical) counterparts.  In future work we
intend both to investigate the relative stabilities of the striped and
N\'{e}el$^{\ast}$ GS phases in the intermediate regime $\kappa_{c_{1}}
< \kappa < \kappa_{c_{2}}$, and to perform a rigorous study of the
stability of possible phases with VBC order, particularly those with
PVBC and CDVBC ordering, in the regime $\kappa > \kappa_{c_{2}}$.

\section*{Acknowledgment} 
We thank the University of Minnesota Supercomputing Institute for
Digital Simulation and Advanced Computation for the grant of
supercomputing facilities, on which we relied heavily for the
numerical calculations reported here.

\end{document}